\def\be{\begin{equation}}
\def\ee{\end{equation}}
\def\bea{\begin{eqnarray}}
\def\eea{\end{eqnarray}}

\def\la{\langle}
\def\ra{\rangle}
\documentclass[english]{elsart}

\usepackage{amssymb}
\usepackage{babel}
\usepackage[dvips]{graphicx} 
\usepackage{listing}
\begin{document}

\begin{frontmatter}



\title{Statistical Characterization of a 1D Random Potential Problem
 -- with applications in score statistics of MS-based  peptide sequencing}


\author{Gelio Alves and Yi-Kuo Yu\protect\footnotemark
}

\address{National Center for Biotechnology Information,
 National Library of Medicine,
 National Institutes of Health, Bethesda, MD 20894  
}

\footnotetext{\scriptsize To whom correspondence should be addressed.  E-mail address:
 yyu@ncbi.nlm.nih.gov
}
\begin{abstract}
We provide a complete thermodynamic solution of a 1D hopping model in
 the presence of a random potential by obtaining the density of states.
 Since the partition function is related to the density of states by a
 Laplace transform, the density of states determines completely the thermodynamic
 behavior of the system. We have also shown that the transfer matrix technique, or the so-called
 dynamic programming, used to obtain the density of states in the 1D hopping model may be generalized to
 tackle a long-standing problem in statistical significance
 assessment for one of the most important {\em proteomic} tasks --
 peptide sequencing using tandem mass spectrometry data. 

\end{abstract}

\end{frontmatter}

\section{Introduction}
\label{intro}
Important in both fundamental science and numerous applications, 
 optimization problems of various degrees of 
complexity are challenging (see ~\cite{Garey} for an 
  excellent introduction). Optimization conditioned by 
 constraints that may vary from event to event is of especial 
 theoretical and practical importance. As a first example, when dealing with 
a system under a random potential, each realization
 of the random potential demands a separate optimization resulting in a different ground
 state.  The thermodynamic behavior of such a system in a quenched random potential
 crucially depends on the random potential realized. A similar but practical 
 problem may arise in routing passengers at various cities to reach their destinations. In the latter case,
 the optimal routing depends on the number of passengers at various locations, the costs from one location to the others,
 which likely to vary from time to time. This type of conditional optimization also occurs in modern proteomics problem,
 that is, in the mass spectrometry (MS) based peptide sequencing. In this case, each tandem MS (MS$^2$) spectrum constitute
  a different condition for optimization which aims to find a database peptide or a {\it de novo} peptide to
 best explain the given MS$^2$ spectrum.  

 When the cost function of an optimization problem can 
 be expressed as a sum of independent local contributions, 
 the problem usually can be solved using the transfer matrix method  
 that is commonly employed in statistical physics. A well-studied example 
 of this sort in statistical physics is the directed polymer/path in a
 random medium (DPRM)~\cite{Huse_85a,Kardar_87,Fisher_91}.  Even when a small non-local energetics is involved, 
 the transfer matrix approach still proves useful~\cite{DAY_05}. As an example, the close relationship between
 the DPRM problem and MS-based peptide sequencing, where a small nonlocal energetics is necessary to
 enhance the peptide identifications, was sketched in an earlier publication~\cite{DAY_05} and the 
 cost value distribution from many possible solutions other than the optimal one is explored. Indeed,
 obtaining the cost value distribution from {\it all} possible solutions in many cases is 
 harder than finding the optimal solution alone. 
 In this paper, we will provide the solution to a generic problem that enables a full characterization of the 
  peptide sequencing score statistics, instead of just the optimal peptide. 
 The 1D problem considered is essentially a hopping model in the
 presence of a random potential. The solution to this problem may also be useful in other applications 
 such as in routing of passengers and even internet traffic. 

In what follows, we will first introduce the generic 1D hopping model in a random potential, followed by 
its transfer matrix (or dynamic programming) solution. We then discuss the utility of this solution
 in the context of MS-based peptide sequencing, and demonstrate with real 
 example from mass spectrum in real MS-based proteomics experiments. In the discussion section, we will
 sketch the utility of the transfer matrix solution in other context and then conclude with 
 a few relevant remarks. 

\section{1D hopping in random potential}
\label{model}
Along the $x$-axis, let us consider a particle that can hop with a set of prescribed distances $\{ m_i \}_{i=1}^K$ 
 towards the positive $\hat x$ direction. That is, if the particle is currently at location $x_0$, it can
 move to location $x_0+m_1$, $x_0+m_2$, \ldots $x_0+m_K$ in the next time step. At each hopping step, the particle 
 will accumulate an energy $-s(x)$ from location $x$ that it just visited. 
 The score $s(x)$ (negative of the on-site potential energy) is assumed positive and 
 may only exist at a limited number of locations. For locations 
 that $s(x)$ do not exist, we simply set $s(x) = 0$ there. 
 The energy of a path starting from the origin 
 specified by the sequential hopping events $ p\equiv \{ m_{h_1}, 
 m_{h_2}, \ldots, m_{h_L} \}$ would have visited locations $\{
x_1, x_2, \ldots, x_L \}$ with $x_i \equiv \sum_{j=1}^i m_{h_j}$ and 
 has energy 
\[
E_p(x = x_L) \equiv - \sum_{i=1}^{L-1} s\left( x_i \right) \equiv -S_p(x)\; .
\] 
In general, there can be more than one path terminated at the same point. Treating each path as 
 a state with energy given by $E_p$, one ends up having the following recursion relation
 for the partition function $Z(x) \equiv \sum_p e^{-\beta E_p(x)}$
\be \label{pf_global.0}
  Z(x) = \sum_{i=1}^K e^{\beta s(x-m_i)} Z(x-m_i) \; ,
\ee
where $\beta = 1/T$ plays the role of inverse temperature (with $k_B = 1$ chosen). 
If one were only interested in the best score terminated at point $x$, it will be given
 by the zero temperature limit $\beta \to \infty$ and the recursion relation may be obtained
 by taking the logarithm on both sides of (\ref{pf_global.0}) and divided by $\beta$ then taking $\beta \to \infty$ limit
 to reach  
\be \label{pf_zt_best}
S_{\rm best}(x) = \max_{1\le i \le K} \{ s(x-m_i)+S_{\rm best}(x-m_i) \}\;,
\ee
where $S_{\rm best}(x)$ records the best path score among all paths reaching position $x$.  
 This update method, also termed dynamic programming, 
 records the lowest energy and lowest energy path 
 reaching a given point $x$. The lowest energy among all
 possible at position $x$ is simply $-S_{\rm best}(x)$ 
 and the associated path can be obtained by tracing backwards 
 the incoming steps. It is interesting to observe that one can also obtain the worst
 score at each position via dynamical programming
\be \label{pf_zt_worst}
S_{\rm worst}(x) = \min_{1\le i \le K} \{ s(x-m_i)+S_{\rm worst}(x-m_i) \}\; .
\ee

The full thermodynamic characterization demands more information than the 
 ground state energy. In principle, one may obtain the full partition function 
 using eq.~(\ref{pf_global.0}) evaluated at various temperatures. This procedure, however,
 hinders analytical property such as determination of the average energy 
\[
\la E \ra \equiv -\frac{\partial \ln Z}{\partial \beta} \;.
\]
A better starting point may be achieved if one can obtain the density of states $D(E)$.
 In this case, we have 
\bea
Z &\equiv& \int dE e^{-\beta E} D(E) \nonumber \\
\la E \ra & = & \frac{\int dE e^{-\beta E} E D(E)}{\int dE e^{-\beta E} D(E)} \; .\nonumber
\eea
Note that if the ground energy $E_{\rm grd}$ of the system is bounded from below, the partition function
 is simply a Laplace transform of a modified density of states given by
\[
Z = e^{-\beta E_{\rm grd}} \int_0^\infty dE e^{-\beta E} \tilde D(E)
\]
where $\tilde D(E) \equiv D(E-E_{\rm grd})$  and 
\[
\la E \ra = E_{\rm grd} + \frac{\int_0^\infty dE e^{-\beta E} E \tilde D(E)}{\int_0^\infty dE e^{-\beta E} \tilde D(E)}
\]
This implies that the density of states $D(E)$ together with the ground state energy $E_{\rm grd}$ determine
 all the thermodynamic behavior of the system. In the next section, we will explain how to obtain the density of
 states using the dynamical programming technique as well as how to extend this approach to more complicated situations that will
 be useful in characterizing the score statistics in MS-based peptide sequencing.

\section{Obtaining the Density of States}
\label{dos}
The density of states is related to the energy histogram in a simple way.
The number of states between energies $E$ and $E+\eta $ (with $\eta \ll 1$) 
is given by $D(E) \eta $. If we happen to use $\eta $ as the energy bin size
 for energy histogram, the count $C(E)$ in the bin with energy $E$ is simply $D(E) \eta$ and
 the density of states $D(E) = C(E)/\eta$. For simplicity,
 we will assume that the all the on-site energies $-s(x)$ are integral multiple of $\eta$.
 This implies that each path energy/score is also an integral multiple of $\eta $.  
  In the following subsections, we will use score density of states instead of energy density of 
 states.

\subsection{The Simplest Case and its Application}
We denote by $C(x,N)$ the number of paths reaching position $x$ with score $N\eta$. 
With this notation, we can easily write down the recursion relation for $C(x,N)$
 as follows
\be \label{s.dos}
C(x,N) = \sum_{i=1}^K C(x-m_i, N-\frac{s(x-m_i)}{\eta}) \; .
\ee 
This recursion relation allows us to compute the density of states
 in the same manner as computing the partition function (\ref{pf_global.0})
 except that we need to have an additional dimension for score at each position $x$.
 As an even simpler application of this recursion relation, suppose that one is only
 interested in the number of paths reaching position $x$, one may sum over the energy part on both side
 of (\ref{s.dos}) and arrives at 
\be \label{p_count}
C(x) = \sum_i C(x-m_i) ,
\ee
which enables a very speedy way to compute the total number of paths reaching position $x$. 
In the context of {\it de novo} peptide sequencing~\cite{Lutefisk}, this number corresponds to the total number
 of {\t all possible} {\it de novo} peptides within a given small mass range. Although simply obtained,
  this number may be useful for providing rough statistical assessment in {\it de novo} 
 peptide sequencing.

\subsection{The More Realistic Case}
In general, one may wish to associate with each hop an energy $h$ or one may wish to introduce
 some kind of score normalization based on the number of hopping steps. This is indeed the case when
 applying this framework to MS-based peptide sequencing where a peptide length factor 
 adding or multiplying to the overall raw score is a common practice. 
In this case, it becomes important to keep track the number of hops made in each path. We may
 further categorize the counter $C(x,N)$ into $\sum_L C(x,N,L)$. That is, we may separate the paths with
 different number of steps from one another and arrive at a finer counter $C(x,N,L)$ which records the 
 number of paths reaching position $x$ with score $N\eta$ and with $L$ hopping steps. 

It is rather easy to write down the recursion relation obeyed by this fine counter
\be \label{sl.dos}
C(x,N,L) = \sum_{i=1}^K C(x-m_i, N-\frac{s(x-m_i)}{\eta},L-1) \; .
\ee
This recursion relation allows us to renormalize the raw score based on the number of steps taken.
 For example, for RAId{\_}DbS~\cite{RAId_DbS}, a database search method we developed, we divide the 
 raw score obtained by $2(L-1)$ for any peptide (path) of $L$ amino acids (hopping steps) to get
 better sensitivity in peptide identification.  

In principle, the recursion relations given by (\ref{s.dos}-\ref{sl.dos}) are all one-dimensional updates.
 The only difference is the internal structure of counters at each position $x$. 
 For (\ref{p_count}), the counter is just an integer and has no further structure. For (\ref{s.dos}), the 
 counter at each position has a 1D structure indexed by the score. For (\ref{sl.dos}), the counter at each
 position $x$ has a 2D structure indexed by both the score and the number of hopping steps. This means that
 in terms of solving the problem using dynamical programming, it is always a 1D dynamical programming 
 with different degrees of internal structure that may lengthen the execution time 
  when shifting from the simplest case (\ref{p_count}) to the more complicated case (\ref{sl.dos}). 
Obviously at each position $x$, there is an upper bound and a lower bound for score and
 also for the number of hopping steps accumulated. 
 We shall call them $S_{\rm best}(x)$, $S_{\rm worst}(x)$,
 $L_{\rm max}(x)$ and $L_{\rm min}(x)$ respectively. The first two quantities may be obtained
 by eqs.~(\ref{pf_zt_best}) and (\ref{pf_zt_worst}) respectively. We provide the recursions for  the 
 two latter quantities below
\bea
L_{\rm max}(x) &=& \max_{1\le i \le K} \{ L_{\rm max}(x-m_i) \} + 1 \label{l_max} \; ,\\
L_{\rm min}(x) &=& \min_{1\le i \le K} \{ L_{\rm min}(x-m_i) \} + 1 \label{l_min} \; .
\eea

Eqs.~(\ref{pf_zt_best}-\ref{pf_zt_worst}) and (\ref{l_max}-\ref{l_min}) provides the ranges for 
 both the scores and the number of cumulative hopping steps at each position $x$ via simple
 dynamic programming. As we will discuss later, this information enables a memory-efficient 
  computations of score histograms.

\section{Application in MS-based Peptide Sequencing}
\label{MSMS}

 In this section, we focus on an important subject in modern
 biology -- using MS data to identify the
  numerous peptides/proteins involved in any given biological process. 
 Because of the peptide mass degeneracies and the limited measurement
 accuracy for the peptide mass-to-charge ratio, using 
 MS$^2$ spectra is more effective in peptide identifications. 
 In a MS$^2$ setup, a selected peptide with
 its mass identified by the first spectrometer is fragmented 
 by noble gas, and the resulting fragments are 
 analyzed by a second mass spectrometer.
 Although such MS$^2$-based proteomics approaches promise high throughput 
 analysis, the confidence level assignment for any peptide/protein
 identified is challenging. 

 The majority of peptide identification methods are
 so-called database search approaches. The main 
 idea is to {\em theoretically} fragment each peptide in a
 database to obtain the corresponding {\em theoretical} spectra.
 One then decides the degree of similarity between each theoretical 
 spectrum and the input query spectrum using a scoring function.
 The candidate peptides from the database
  are ranked/chosen according to their similarity 
 scores to the query spectrum. Although one may assign relative 
 confidence levels among the candidate peptides via various (empirical) 
 means, an objective, standardized calibration exists only recently~\cite{E_calib}.   
  In our earlier publications~\cite{DAY_05,RAId}, 
 we proposed to tackle this difficulty by using a {\em de novo} 
 sequencing method to provide an objective 
 confidence measure  that is both database-independent and takes into account 
  spectrum-specific noise. In this paper, we will provide concrete algorithms 
 for such purpose. 

 To begin, consider a spectrum $\sigma$ with parent ion mass range
 $[w - \delta, w + \delta]$,
  we denote by $\Pi(w,\delta)$ the set of all
  ``possible'' peptides with masses  in this range.
 Given a peptide $\pi$ from $\Pi(w,\delta)$, the associated
 quality score $S(\pi,\sigma)$ is defined by a {\em prescribed}
  scoring system.  
  The score distribution of $S(\pi,\sigma)$
 within $\Pi(w,\delta)$ provides naturally a likelihood measure
 for any given peptide $\pi$ to the the correct one. 

However, as described earlier~\cite{DAY_05}, this seemingly straightforward idea 
  faces two difficulties in terms of implementations.
 First, unlike the DPRM problem for which the  function to be
 optimized is defined without ambiguity,  the choice of
  the scoring function is somewhat empirical because the
  parameters used in the scoring must be trained using a training data set.
 Further, because of different instruments and experimental setups,
 it seems impossible to design a scoring system such that the correct
 peptide for each spectrum has the highest score among {\em all possible} peptides;
 the application of a given scoring function to general
 cases may require a leap of faith. Second,  even after the scoring function is chosen,
 it is not known how to find the peptide $\pi_o$ that maximizes $S(\pi, \sigma)$ as well as
 the score distribution ${\rm pdf}(S)$ within $\Pi(w,\delta)$
 other than by the generally impractical
 procedure of examining all members of $\Pi(w,\delta)$. 

The first difficulty can be alleviated by validating
  high scoring {\em de novo} peptides via database searches~\cite{RAId}
 and is not the main focus of the current paper. Note that a partial solution 
 to the second problem via iterative mapping when nonlocal score 
 contributions exist is provided earlier~\cite{DAY_05}. Here we tackle the second problem head
 on when the scoring function used does not contain nonlocal contribution other than a  
 final renormalization with respect to the peptide length. Our algorithms contains two
 parts: computer memory allocations and dynamical programming update. Prior to discussing 
 these two parts, however, we first address the important issue of choosing a
 good mass unit.    

\subsection{Choosing a Good Mass Unit}
The goal here is to choose a mass unit $\Delta$ and expresses the molecular mass of each amino acid
 as an integral multiple of this unit. For example, one may choose $\Delta$ to be $0.1$ Dalton (Da),
 and round the molecular mass of each amino acid to be an integral multiple of $0.1$ Da.
 Once a mass unit is chosen, all the masses under consideration are integral multiples of this 
 unit. It turns out that different choices of the mass unit leads to different maximum cumulative mass error.
 As a specific example, consider using $\Delta = 0.1$ Da as the mass unit. The mass of 
 Alanine, with true mass $71.03711538$ Da, is now represented as $710 \Delta$. This molecular mass expression 
 is $0.03711538$ Da smaller than the true molecular mass of Alanine. When this happens, the integral
 mass representation has a mass smaller than the true mass, and we call such type of mass error
 a {\it down}-error. Now the amino acid Tryptophan with molecular mass $186.07931613$ Da 
 will be assigned an integral mass of $1861 \Delta$, which has an extra of
 $(0.1-0.07931613)$ Da compared to the true mass. We call this type of mass error the {\it up}-error. 

The ratio of the mass error to the real molecular mass when multiplied by $3000$ Da
  provides the cumulative maximum error that can be induced by a single amino acid
  at $3,000$ Da mass. For a fixed mass unit, we went over this mass error analysis for
 each of the twenty amino acids and documented the largest up-error and down-error.
 The larger one between the maximum up-error and the maximum down-error is called the max-error. 
  To search for best mass units that minimize the max-error at $3,000$ Da, we went over all possible mass 
 unit ranging from $0.005$ Da to $1.005$ Da in step of $10^{-6}$ Da. Interestingly enough, we found 
 a discrete list of mass units that have smaller max-error compared to their nearby 
 mass units. These numerically found {\it magic} mass units are summarized in table~\ref{tab.magic}. 

\begin{table*}
\caption[]{A list of best mass units in Da. The abbreviation ``m.u.e." stands for 
 ``maximum up-error," while ``m.d.e." stands for ``maximum down-error." 
The maximum up-error, maximum down-error, and max-error are 
 evaluated in extrapolation to $3,000$ Da as described in the text. The abbreviation ``a.a.w.m.u." 
 stands for ``amino acid with maximum up-error," while ``a.a.w.m.d." stands for ``amino acid with maximum down-error."
 \vspace*{10pt}} \label{tab.magic}
\begin{tabular}{cccccc}\hline
mass unit &        m.u.e. &     a.a.w.m.u. &   m.d.e. &    a.a.w.m.d.    &  max-error \\ \hline
0.006070  &       0.041980 &       Tryptophan &     0.037455 &    Cysteine    &           0.041980 \\
0.007300  &       0.041495  &      Methionine &     0.061276  &   Asparagine     &        0.061276 \\
0.017540  &       0.094183    &    Cysteine   &   0.121977     &   Proline     &         0.121977\\
0.021500  &       0.199585    &    Arginine  &    0.182283    &   Asparagine      &          0.199585\\
0.054470    &     0.453793     &   Asparagine  &    0.347792    &  Alanine       &          0.453793\\
0.065400    &     0.553492     &   Lysine  &   0.536989     &   Alanine      &        0.553492\\
0.109450    &     0.908287     &    Proline  &   0.900898     &   Lysine     &        0.908287\\
0.110300    &     0.962781     &  Histidine  &   0.858742     &   Lysine     &        0.962781\\
0.110320    &     0.960176     &   Aspartate  &    0.907801    &   Histidine    &          0.960176\\
0.500208    &     0.980357     &   Cysteine  &   0.983149     &    (Iso)Leucine    &        0.983149\\
1.000416    &     0.980357     &   Cysteine  &    0.983149    &    (Iso)Leucine    &          0.983149 \\ \hline
\end{tabular}
\vspace*{0.2in}
\end{table*}

 Once a mass unit is chosen, all the amino acid masses are effectively integers. 
 To obtain the score histogram of {\it all} {\it de novo} peptides when queried
 by a spectrum $\sigma$ with parent molecular mass $w$ (with N- and C- terminal 
 groups of the peptide stripped away), we first construct a mass array where index $k$ corresponds
 a molecular mass $k\Delta$. To encode all possible peptides with molecular mass
 up to $w$, we need to have an array of size $w/\Delta + 1$. Apparently, when a 
 larger mass unit is used, the size of the mass array is smaller and thus 
 reduces computation time. However, as one may see from table~\ref{tab.magic}, 
 the larger mass unit is also accompanied by a larger max-error and might not be preferred
 when high mass accuracy is the first priority. 

\subsection{Efficient Memory Allocation}
The basic idea of our algorithm is to encode all possible peptides
 in the mass array by linking pointers, analogous to the consecutive hopping
 steps in the 1D hopping model. For an amino acid $a$, let $n(a)$ represents its 
 corresponding integer mass in unit of $\Delta$. For a peptide made of $[a_1, a_2,\ldots, a_M]$, 
 it will have a hopping trajectory in the molecular array given by $[0,x_1, x_2,\ldots, x_M]$ 
 with $x_{i\ge 1} \equiv \sum_{j=1}^i n(a_j)$. Let us also denote $x_M$ by $x_F$ to indicate that it is the 
 terminating point of the path. Apparently, all possible peptides with molecular masses equal to   
 $x_F \Delta$ will all have corresponding hopping paths starting at the origin and terminating at $x_F$. 
 Through appropriate pointer linking, one may therefore encode {\it all} possible 
 peptides with molecular mass $x_F \Delta$ in a one-dimensional mass array.

For a given spectrum $\sigma$, depending on the score function used, one may calculate local score contributions 
at each mass index. This step is done once only for the whole mass array, and need not be repeated for each candidate peptide.
In a typical MS$^2$ experimental spectrum, there always exists some level of 
 parent ion mass uncertainty. Once the size of the mass uncertainty is specified, 
 we only need to examine {\it de novo} peptides whose corresponding
 hopping paths terminating at a few consecutive mass indices. This indicates that some of the mass indices of
 the aforementioned mass array may not even be used in this context. Below we describe how to efficiently obtain 
{\it relevant mass indices} and only allocate computer memories for those masses.   

Assume that the possible terminating points are $F_1, F_2, \ldots, F_k$ with $F_{j+1} = F_j + 1$.
The update rules described in Eqs.~(\ref{pf_zt_best}-\ref{pf_zt_worst}), (\ref{p_count}), and (\ref{l_max}-\ref{l_min})
 will also be used at this stage. The following pseudocode describes our algorithm. 

{\small \tt 
\begin{listing}
Initialize the mass\_index = 0 entry\\
$S_{\rm best} = S_{\rm worst} = L_{\rm max} = L_{\rm min} = 0$; $C$=1;\\
REMARK: Max\_aa is the maximum number of amino acids considered \\
for (aa\_index = 0; aa\_index $<$ Max\_aa; aa\_index ++) \{\\
\hspace*{12pt} label occupancy of n(aa\_index);	\\
\hspace*{12pt} at n(aa\_index) attach a pointer back to 0;	\\
\hspace*{12pt} update $S_{\rm best}$, $S_{\rm worst}$, $L_{\rm max}$, $ L_{\rm min}$, $C$ at  n(aa\_index);\\
\} \\
for (mass\_index = 1; mass\_index $<= F_k$; mass\_index ++)\{\\
\hspace*{12pt} if (mass\_index occupied ?) \{\\
\hspace*{24pt} for (aa\_index = 0; aa\_index $<$ Max\_aa; aa\_index ++) \{\\
\hspace*{36pt} label occupancy of (mass\_index + n(aa\_index));	\\
\hspace*{36pt} at mass\_index+n(aa\_index) attach a pointer to mass\_index; \\
\hspace*{36pt} update $S_{\rm best}$, $S_{\rm worst}$, $L_{\rm max}$, $ L_{\rm min}$, $C$ at (mass\_index + n(aa\_index));\\
\hspace*{24pt} \}\\
\hspace*{12pt} \}\\
\}\\
for (mass\_index = $F_k$ ; mass\_index $ >= F_1$; mass\_index --)\{\\
\hspace*{12pt} backtrack all possible paths $\to$ final occupied entries;\\
\}\\
\end{listing}
}

The last step in the algorithm above identifies {\it relevant mass indices}, mass\_indices that will be traversed by 
the hopping paths 
 of all peptides with molecular masses in the range $[F_1\Delta, F_k\Delta]$. We only need to allocate computer memory 
 associated with those sites. For each of these relevant sites, we also know the values of $S_{\rm best}$, $S_{\rm worst}$,
 $L_{\rm max}$, $L_{\rm min}$, and the total number of peptides reaching that site through the algorithm above. 
 One may therefore allocate a 2D array of size $(S_{\rm best}(i) - S_{\rm worst}(i))/\eta \times (L_{\rm max}(i) - L_{\rm min}(i))$ 
 for each relevant mass\_index $i$ for later use. 
 
\subsection{Main Algorithm and some Results} 
 
Once memory allocation for relevant mass\_indices is done, we can efficiently go through those relevant sites to obtain the 
 2D score histogram that we mentioned. In the pseudocode below, update is performed using eq.~(\ref{sl.dos}). 
 We now demonstrate the very simple main algorithm

{\small \tt 
\begin{listing}
Initialize all the fine counters $C(x,N,L)= 0 $\\
except $C(x=0,N=0,L=0)$=1;\\
for (aa\_index = 0; aa\_index $<$ Max\_aa; aa\_index ++) \{\\
\hspace*{12pt} update $C(x,N,L)$ at $x=$n(aa\_index);\\
\} \\
for (mass\_index in ascendingly ordered {\it relevant mass\_indices})\{\\
\hspace*{12pt} for (aa\_index = 0; aa\_index $<$ Max\_aa; aa\_index ++) \{\\
\hspace*{24pt} update $C(x,N,L)$ at $x = $(mass\_index + n(aa\_index));\\
\hspace*{12pt} \}\\
\}\\
\end{listing}
}
 
We now define the final 2D counter
\be \label{2Dhist_final}
Y(N,L) \equiv \sum_{i=1}^k C(F_i, N, L) \; .
\ee 
Apparently, in the 1D hopping model when allowing $k$ consecutive terminating points, 
 the resulting density of states $D(E)$ can now be expressed as $D(E = - N\eta) = \sum_L Y(N,L)/\eta$.
If one were interested in normalizing the final score in a path-length dependent manner,
 one will has the following generic transformation
\be
H(E) = \sum_{L}\int dE' \frac{Y(E'= - N\eta,L)}{\eta} \delta \left(E-f(E',L)\right)
\ee
where $f(E',L)$ is a generic length-normalized energetic function that takes the raw energy $E'$ with $L$ hopping
 steps and turn them into a new energy $f(E',L)$, and $\int dE' \to  \eta \sum_N $ is understood.

\begin{figure}
\begin{center}
\includegraphics[width= 0.6\columnwidth,angle=-90]{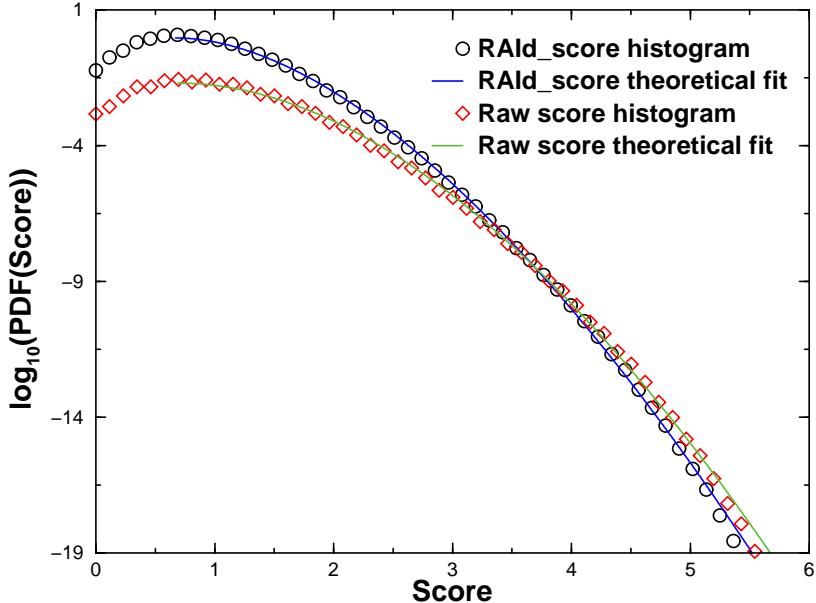}
\vspace*{-0.1in}
\end{center}
\caption[]{Score histograms for raw score and RAId\_DbS score. Note that the two histogram cross each other at
 large score regime, indicating  that the raw score function might not be as effective as the RAId\_DbS score,
 see text for details.}
\label{fig1}
\end{figure}

Using a real experimental MS$^2$ spectrum of parent ion mass $2254.7 \pm 3.0$ Da and a raw scoring 
 function (RAId\_DbS~\cite{RAId_DbS} raw score without divided by $2(L-1)$ with $L$ being the peptide length),
 we obtained a 2D score histogram. From this 2D score histogram, we can compute the average peptide length $\la L \ra$
 as well. We then transform the 2D score histogram using two different $f$ functions. In the first case,
 $f(E',L) = E'/2(\la L \ra -1)$, meaning that one just divides the score by a constant given by $2(\la L \ra - 1)$.  
 In the second case, we use the RAId\_DbS scoring function where $f(E',L) = E'/2(L-1)$.  
In Figure~\ref{fig1}, 
  we show the two resulting score histograms along with the fits to theoretical distribution function~\cite{RAId_DbS}.  
 As one may see from the figure, both histograms are well fitted by the theoretical distribution function
 over at least 15 order of magnitudes. There is difference, however, in the histograms obtained.
 In the first case, where the score is merely divided by the average length, we have a wider score distribution than
 that of the second case. This implies that a high scoring hit out of the  first type of scoring function will have
  a larger $P$-value than that of the second type. This is  perfectly reasonable because when using
  the first type of raw scoring, very long peptides which by random chances are more likely to hit on 
  fragment peaks in the mass spectrum are less penalized than the shorter peptides. As a consequence, one anticipates
 more false long peptides out of the first type of scoring method than that of the second scoring method.
 Therefore, one should assign a larger $P$-value to the former case and a smaller $P$-value to the latter case.
 It is apparently important to be able to obtain score histograms of the second scoring method. However, this can
 only be achieved if one keeps the length information in the dynamical programming update, see eq.~(\ref{sl.dos}).    

\section{Discussion, Summary, and Outlook}
\label{summary}
Our method may also be extended to other applications. In the case of passenger routings,
 the $x$-axis actually represents time. The local score may be viewed as the additional cost 
 that may vary for different stops. Once the problem is laid out, the 2D histogram obtained from
 our solution indicates the number of equivalent routes in terms of additional costs and the
 total number of stops. This problem should be interesting in its own right.

In this paper, we developed a new approach to obtain the density of states of a 1D hopping problem in
 random potential. We have extended the simplest case scenario and
 have shown that we can apply this method to provide a {\it complete} score histogram
 for MS-based peptide sequencing problem. This important information may be used for 
 a more objective statistical significance assignment in peptide identification. 
 Our algorithm may also serve as a speedy {\it de novo} algorithm. If one is only interested in
 getting the best scoring peptide with length normalized score, one only needs to keep track of 
$S_{\rm best}(x,L)$. Furthermore, it is straightforward to include in our {\it de novo} algorithm
 post-translationally modified amino acids. The effect is simply an enlargement of the alphabet.
 That is, instead of having 20 amino acids, we will simply have more allowed masses but without needing 
 to change any part of the algorithm.     

In the near future, we would like to build a web application that allows the users to 
 obtain information of interest. For example, a user might be interested in knowing:  
 given a parent ion molecular mass and a mass error tolerance, how many {\it de novo}
 peptides can there be? Furthermore, we plan to provide users with the full
 score histogram when a query spectrum is provided and a scoring method is chosen. 
Our approach, founded on statistical physics, can easily 
 address this type of questions to provide useful information for biological researches.

\section*{Acknowledgement}
This work was supported by the Intramural
Research Program of the National Library of Medicine at the National Institutes of Health.


\begin{thebibliography}{00}

\bibitem{Garey} M.R. Garey and D.S. Johnson,
 {\em Computers and intractability}, W.H. Freeman and company, New York
 (1979). 

\bibitem{Huse_85a} D.A. Huse and C.L. Henley, Phys. Rev. Lett. {\bf 54}, 2708 
 (1985).

\bibitem{Kardar_87}
 M. Kardar, Nucl. Phys. {\bf B290}, 582 (1987).

\bibitem{Fisher_91} 
D.S. Fisher and 
 D.A. Huse, Phys. Rev. B. {\bf 43}, 10728 (1991).

\bibitem{DAY_05}
T.P. Doerr, G. Alves and Y.-K. Yu, Physica A
{\bf 354}, 558-570 (2005).

\bibitem{Lutefisk}
J.A. Taylor and R.S. Johnson,  
 {\it Raid Commu. Mass Spect.}  {\bf 11}, 1067 (1997) .  


\bibitem{RAId_DbS}
G. Alves, A.Y. Ogurtsov and Y.-K. Yu, 
  Biology Direct {\bf 2}, art. no. 25 (2007).

\bibitem{E_calib}
G. Alves, A.Y. Ogurtsov, W.W. Wu, G. Wang,
R.-F. Shen  and Y.-K. Yu,
 Biology Direct {\bf 2}, art. no. 26 (2007).

\bibitem{RAId}
G. Alves and Y.-K. Yu, 
 Bioinformatics {\bf 21}, 3726-3732 (2005).






\end{thebibliography}
\end{document}